\documentclass[fleqn,twoside]{article}
\usepackage{espcrc2}

\usepackage{amssymb}
\usepackage{url}

\usepackage{graphicx}
\usepackage[figuresright]{rotating}

\newcommand{\AmS}{{\protect\the\textfont2
  A\kern-.1667em\lower.5ex\hbox{M}\kern-.125emS}}

\title{RooStatsCms: a tool for analyses modelling, combination and statistical 
       studies.}

\author{D.~Piparo\address[UNIKA]{Institut f\"ur Experimentelle Kernphysik, 
                                 Universit\"at Karlsruhe},
        G.~Schott\addressmark[UNIKA]
        and
        G.~Quast\addressmark[UNIKA]}

\begin{document}

\begin{abstract}
The RooStatsCms (RSC) software framework allows 
analysis modelling and combination, statistical studies together with 
the access to sophisticated graphics routines for results visualisation.
The goal of the project is to complement the existing analyses by 
means of their combination and accurate statistical studies.
\vspace{1pc}
\end{abstract}
%
\maketitle
%
%
Soon the LHC machine will open a new \mbox{exciting} 
era of measurement campaigns. 
A reliable and widely accepted tool for analyses combination and statistical 
studies is needed at first for limits estimations. Then it should provide the 
basis for discoveries and finally parameter measurements. Our proposed 
solution is RooStatsCms \cite{RooStatsCms}.
%
%
\section{Statistical methods}
\label{statistical_methods}
\subsection{Separation of signal plus background and background only hypotheses}
\label{hypotheses_separation}
\begin{figure}[h]
    \begin{center}
        \label{m2lnQ}
        \includegraphics[width=70mm]{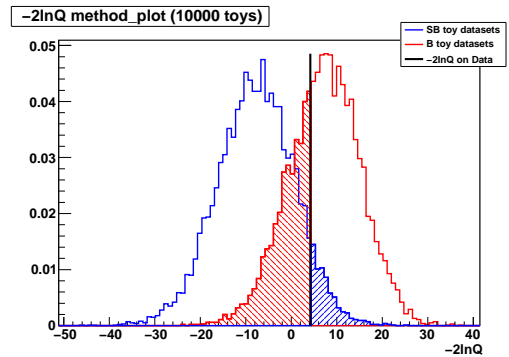}
        \caption{The distributions of $-2\ln Q$ in the background only (red, 
                 on the right) and signal plus background (blue, on the left) 
                 hypotheses. The black line 
                 represents the value of $-2\ln Q$ on the measured data. 
                 The shaded areas represent $1-CL_{b}$ (red) and $CL_{sb}$ 
                 (blue). The $-2\ln Q$ variable can be used instead of $Q$ 
                 to ease the calculations without any loss of generality.}
    \end{center}
\end{figure}

The analysis of search results can be formulated in terms of a hypothesis 
test. The first step consists in identifying the observables which comprise 
the results, e.g. the number of signal events observed or the 
cross section for a particular process.
Then a test statistic is specified. A good choice is 
$Q'=-2\ln Q$, where $Q=L_1/L_0$ and $L_0$ and $L_1$ are, for a given dataset, 
the likelihoods computed in the $\mathcal{H}_0$ (background only, b) and 
$\mathcal{H}_{1}$ (signal plus background, sb) hypotheses respectively.
The last step is to define rules for discovery and exclusion.
In \cite{alexread} is proposed the $CL_s=CL_{sb}/CL_b$ quantity value, 
where 
\begin{equation}
    CL_{sb}=P_{sb}(Q\leq Q_{obs})=\int^{Q_{obs}}_{-\infty}\frac{dP_{sb}}{dQ}dQ 
\end{equation}
and 
\begin{equation}
CL_{b}=P_{b}(Q\leq Q_{obs})=\int^{Q_{obs}}_{-\infty}\frac{dP_{b}}{dQ}dQ.
\end{equation}
The quantity $\frac{dP_{sb}}{dQ}$ is the probability density function 
of the test statistic in the sb experiments (see Fig. \ref{m2lnQ}).
The 95\% confidence level (CL) exclusion of the signal hypothesis is usually 
quoted in case of a $CL_s \leqslant 5\%$.
%
%
\subsection{Profile likelihood (PL)}
\label{profile_likelihood}
\begin{figure}[h]
    \begin{center}
        \label{plscan}
        \includegraphics[width=70mm]{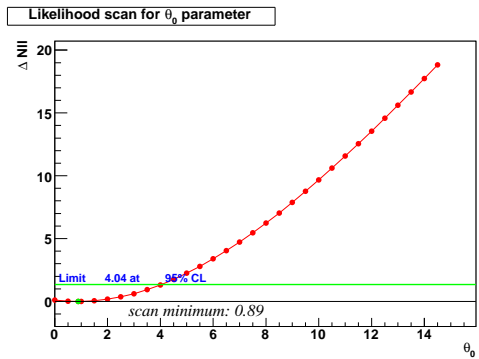}
        \caption{The Likelihood scan in terms of $\Delta nll$ and the 
                 horizontal line for the 95\% CL upper limit 
                 ($\Delta nll \simeq 1.36$). 
                 The interpolation between scan point pairs is linear, 
                 while the minimum of the scan is the minimum of a 
                 parabola built on the lowest three points.}
    \end{center}
\end{figure}

Suppose to have the Likelihood function 
$L(\underline{x},\underline{\theta})$, 
where $\underline{x}$ are the quantities measured for each event and
$\underline{\theta}$ the $K$ parameters of the joint 
pdf describing the data \cite{fredjames}.
The idea of the PL technique
is to select a parameter $\theta_{0}$ and perform a scan over 
a sensible range of values. For \mbox{every} point of the scan, the value of 
$\theta_{0}$ is fixed and the quantity $-\ln L$ ($nll$)
is maximised with respect to the remaining $K$-1 $\theta_{i}$ parameters.
The values are then plotted versus $\theta_{0}$ (Fig. \ref{plscan}). 
Without any loss of generality, the negative loglikelihood 
scan is usually considered in terms of 
$\Delta nll$, with its minimum value equal to 0: this is always possible 
with a shift of all the scan points. 
This curve minimum point abscissa represents the maximum 
likelihood estimator for $\theta_0$, $\hat{\theta}_0$.
Finally, upper limits and intervals of a given 
CL can be read simply from the PL curve, tracing 
horizontal lines at predefined vertical values \cite{Metzger}.
%
%
\section{The RSC framework}
\label{the_rsc_framework}
Based on the RooFit \cite{roofit} technology, RSC is composed of three parts: 
the analysis modelling and combination, 
the implementation of statistical methods and
sophisticated graphics routines for results visualisation.
This framework is intended to be a product for the CMS collaboration but a 
public version, which does not include the modelling part, is available 
under the name of RooStatsKarlsruhe \cite{RooStatsKarlsruhe}.

All the RSC classes inherit from the ROOT \cite{root} TObject. In this way it 
is possible to prototype very quickly macros to be run interactively and all 
the RSC objects can be written on disk in rootfiles exploiting the persistency 
mechanism implemented by ROOT.

The analyses and their combinations are described in ASCII configuration files 
in the ``ini'' format: the \textit{datacards}.
The user at first specifies the observable(s) involved in the analysis.
Thus the signal and background components relative to the observable(s) 
are characterised: the framework also allows to consider multiple 
components for one single observable. 
The single components can be the result of a counting analysis or be 
represented by shapes. Both cases can be taken into account and in presence of 
a shape, the user can decide to describe it through a parametrisation or 
to read it directly from a ROOT histogram stored in a rootfile.
The yields of the components can be described as a single quantity or as a 
product of many factors, 
e.g. $yield = \mathcal{L}\cdot\sigma_{production}\cdot BR \cdot \epsilon$: 
the product of the integrated luminosity, the production cross section, the 
branching ratio and the cuts efficiency.
The systematics on all the the parameters present in the model description 
together with their correlations can also be described in the datacard.

The construction of the RooFit objects representing the (combined) 
analyses is performed through the parsing and processing of the 
\mbox{datacard}. This approach has two advantages: it factorises the analysis 
description from the actual statistical studies and provides to 
the analyses groups a common base to share their results.

The calculations implied by certain statistical methods may be very 
CPU-intensive, e.g.
if the execution of many Monte Carlo toy experiments is foreseen.
The RSC classes that implement those methods are designed to 
ease the creation of multiple jobs intended for a batch system or the grid. 
The outcome is collected in results objects which can be written on disk: 
to every statistical method class corresponds a results class.
Such objects can be recollected and added together therewith profiting 
from a very high Monte Carlo toy experiments statistic.

Special care was devoted to the display of the results: for every results 
class, a plot class is implemented. Therefore in the final step of 
a \mbox{statistical} study the user can get a plot object out of a results 
object and draw it. These two last operations are straightforward and 
need only two lines of \texttt{C++} code to be carried out. 
In addition to the plots directly linked to the results, 
RSC gives the possibility to build other widely accepted plots (see Fig. 
\ref{httexclusion}) with standalone classes.
Examples of plots produced with RSC are Fig. \ref{m2lnQ}, Fig. \ref{plscan}, 
Fig. \ref{httexclusion}.

A big effort was dedicated to the RSC \mbox{documentation} and examples. 
All the classes, methods, members and namespaces are provided with Doxygen 
style comments and a website of RooStatsCms is available \cite{RooStatsCms}.
A wikipage is also present in the official CMS wiki
\footnote{\url{https://twiki.cern.ch/twiki/bin/view/CMS/HiggsWGRooStatsCms}}.
The framework is distributed with example ROOT macros, 
\texttt{C++} programs ready to be compiled and 
Python scripts to ease the creation of the datacards.

A possible application of the RSC implementation of the statistical 
methods described in \ref{hypotheses_separation} 
and \ref{profile_likelihood} is discussed in section \ref{results}.
%
%
\section{A concrete application}
\label{results}
\begin{figure}[h]
    \begin{center}
        \label{httexclusion}
        \includegraphics[width=75mm]{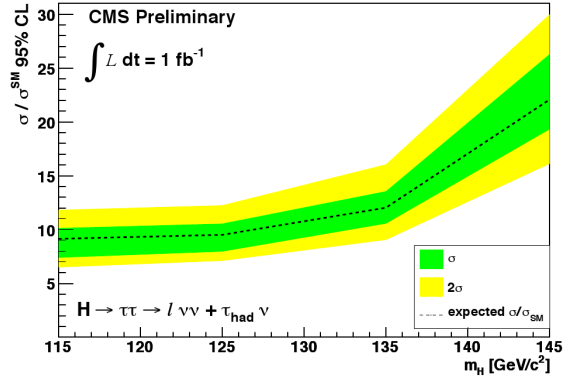}
        \caption{Exclusion plot for the VBF $H \rightarrow \tau \tau$ 
                 CMS analysis. The 1$\sigma$ and 2$\sigma$ bands are obtained 
                 assuming to observe a 1$\sigma$ or 2$\sigma$ upwards 
                 (downwards) fluctuation of the number of observed background 
                 events. The plot is produced with a graphics class of RSC.}
    \end{center}
\end{figure}

For the the CMS VBF $H \rightarrow \tau \tau$ analysis ($1~fb^{-1}$) 
\cite{htautau} study, RSC was used. 
Given the disadvantageous $s/b$ ratio in the various mass hypotheses 
with this integrated luminosity, we put an exclusion limit on the H 
boson production cross section using the technique described in section 
\ref{hypotheses_separation}. Fig. \ref{httexclusion} shows the 
expected $H$ production cross section ($\sigma$) that could be excluded 
with the data available, i.e. the Monte Carlo (MC) 
statistic available after all the cuts, 
in units of Standard Model (SM) $\sigma$. 
To derive the shape of the $\frac{dP}{dQ}$ pdf for the two hypotheses 
several toy MC experiments were performed. 
To take into account the systematics, 
before the generation of each toy MC dataset, the 
parameters values affected by systematic uncertainties were fluctuated 
according to their expected distribution, assumed in this case Gaussian.

\begin{figure}[ht]
    \centering
    \includegraphics[height=70mm , trim= 1mm 1mm 1mm 1mm, clip]{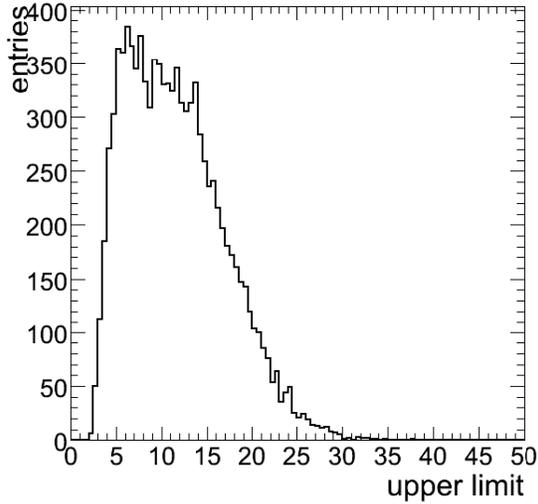}
    \caption{The distribution of the 95\% CL upper limits on the number of 
                signal events for $m_{H}=145$ GeV. The median is located at 
                $10.71$ signal events which corresponds to a 
                $\sigma/\sigma_{SM}$ equal to $6.7$.}
    \label{ULsdistr}
\end{figure}

\begin{figure}
    \centering
    \includegraphics[height=70mm , trim= 1mm 1mm 1mm 1mm, clip ]{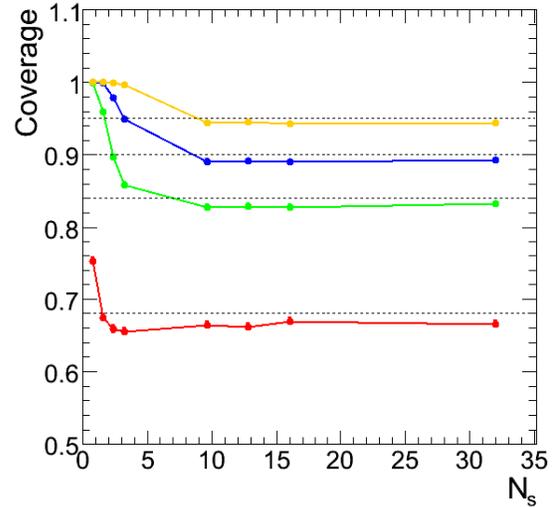}
    \caption{The coverage in case of 68\%, 84\%, 90\%, 95\% CL 
                (coloured lines) versus nominal $N_s$. The reference 
                coverages are shown with the dashed lines.}
    \label{coverage}
\end{figure}

Performing several toy MC experiments we derived also the 
distribution of the upper limits at 95\% confidence level 
shown in Fig. \ref{ULsdistr} 
focusing on the number of signal events ($N_s$). 
In this second study systematic uncertainties were not considered.
An analysis of the coverage, i.e. the fraction of cases in which 
the upper limit is indeed greater than the nominal $N_s$, 
was carried out (Fig. \ref{coverage}). 
The $N_s$ value was also artificially incremented in order 
to study the coverage at different working points. 
Overcoverage is present for low signal yields 
leading to conservative upper limits. 
This feature can be seen as a consequence of the Cram\'{e}r-Fr\'{e}chet 
inequality \cite{Metzger}.
%
%
\section{Conclusions}
\label{conclusions}
Two statistical analyses were carried out using the RSC framework. 
The hypotheses separation study shows that it is possible 
to compute exclusion limits in the context of a hypothesis separation 
test by means of the integrals of the discriminating variables pdf. 
The PL method can be used as well for extracting upper limits 
but it was verified that a significant overcoverage occurs in case of very 
small quantities.

The RooStatsCms proved to be a reliable and valuable tool for 
analyses modelling and complex statistical procedures deployment in a 
simple and flexible way. The graphical routines embedded in the package 
serve as a powerful tool for results visualisation.
%
%

%
%

\begin{thebibliography}{00}
%
%
\bibitem{RooStatsCms} D. Piparo, G. Schott, G. Quast ``RooStatsCms: a tool for analyses modelling, combination and statistical studies'', CMS note (in preparation), \url{www-ekp.physik.uni-karlsruhe.de/~RooStatsCms}.
\bibitem{alexread} A.L. Read ``Modified frequentist analysis of search results (The $CL_{s}$ method)'', CERN open-2000-205.
\bibitem{fredjames} F. James ``Statistical in Experimental Physics 2nd Edition'', Word Scientific 2006.
\bibitem{Metzger} W.J. Metzger ``Statistical Methods in Data Analysis'', Katholieke Universiteit Nijmegen, 2002.
\bibitem{roofit} W. Verkerke, D. Kirkby ``The RooFit toolkit for data Modelling'', \url{roofit.sourceforge.net}.
\bibitem{RooStatsKarlsruhe} D. Piparo, G. Schott, G. Quast ``RooStatsKarlsruhe'',
     \url{www-ekp.physik.uni-karlsruhe.de/~RooStatsKarlsruhe}
\bibitem{root} ROOT: An Object Oriented Data Analysis Framework, \url{root.cern.ch}.
\bibitem{htautau} CMS Collaboration, CMS PAS HIG 08-008.
%
%
\end{thebibliography}
\end{document}